\documentclass[conference]{IEEEtran}
\IEEEoverridecommandlockouts

\usepackage{cite}
\usepackage{amsmath,amssymb,amsfonts}
\usepackage{algorithmic}
\usepackage{graphicx}
\usepackage{textcomp}
\usepackage{xcolor}
\usepackage{enumitem}

\DeclareMathOperator*{\argmin}{arg\,min}

\usepackage{multirow}
\def\BibTeX{{\rm B\kern-.05em{\sc i\kern-.025em b}\kern-.08em
    T\kern-.1667em\lower.7ex\hbox{E}\kern-.125emX}}
\begin{document}

\title{DistDNAS: Search Efficient Feature Interactions within 2 Hours
}

\author{\IEEEauthorblockN{Tunhou Zhang}
\IEEEauthorblockA{\textit{Duke University} \\
Durham, USA \\
tunhou.zhang@duke.edu}
\and
\IEEEauthorblockN{Wei Wen}
\IEEEauthorblockA{\textit{Meta AI} \\
Menlo Park, USA \\
wewen@meta.com}
\and
\IEEEauthorblockN{Igor Fedorov}
\IEEEauthorblockA{\textit{Meta AI} \\
Menlo Park, USA \\
ifedorov@meta.com}
\and
\IEEEauthorblockN{Xi Liu}
\IEEEauthorblockA{\textit{Meta AI} \\
Menlo Park, USA \\
xliu1@meta.com}
\and
\IEEEauthorblockN{Buyun Zhang}
\IEEEauthorblockA{\textit{Meta AI} \\
Menlo Park, USA \\
buyunz@meta.com}
\and
\IEEEauthorblockN{Fangqiu Han}
\IEEEauthorblockA{\textit{Meta AI} \\
Menlo Park, USA \\
fhan@meta.com}
\and
\IEEEauthorblockN{Wen-Yen Chen}
\IEEEauthorblockA{\textit{Meta AI} \\
Menlo Park, USA \\
wychen@meta.com}
\and
\IEEEauthorblockN{Yiping Han}
\IEEEauthorblockA{\textit{Meta AI} \\
Menlo Park, USA \\
yipinghan@meta.com}
\and
\IEEEauthorblockN{Feng Yan}
\IEEEauthorblockA{\textit{University of Houston} \\
Houston, USA \\
fyan5@central.uh.edu}
\and
\IEEEauthorblockN{Hai Li}
\IEEEauthorblockA{\textit{Duke University} \\
Durham, USA \\
hai.li@duke.edu}
\and
\IEEEauthorblockN{Yiran Chen}
\IEEEauthorblockA{\textit{Duke University} \\
Durham, USA \\
yiran.chen@duke.edu}
}

\maketitle

\begin{abstract}
Search efficiency and serving efficiency are two major axes in building feature interactions and expediting the model development process in recommender systems.
Searching for the optimal feature interaction design on large-scale benchmarks requires extensive cost due to the sequential workflow on the large volume of data.
In addition, fusing interactions of various sources, orders, and mathematical operations introduces potential conflicts and additional redundancy toward recommender models, leading to sub-optimal trade-offs in performance and serving cost. 
This paper presents DistDNAS as a neat solution to brew swift and efficient feature interaction design.
DistDNAS proposes a supernet incorporating interaction modules of varying orders and types as a search space. 
To optimize search efficiency, DistDNAS distributes the search and aggregates the choice of optimal interaction modules on varying data dates, achieving a speed-up of over 25$\times$ and reducing the search cost from 2 days to 2 hours.
To optimize serving efficiency, DistDNAS introduces a differentiable cost-aware loss to penalize the selection of redundant interaction modules, enhancing the efficiency of discovered feature interactions in serving.
We extensively evaluate the best models crafted by DistDNAS on a 1TB Criteo Terabyte dataset.
Experimental evaluations demonstrate 0.001 AUC improvement and 60\% FLOPs saving over current state-of-the-art CTR models.
\end{abstract}


\section{Introduction}
Recommender models vary in depth, width, interaction types and selection of dense/sparse features. These versatile feature interactions exhibit different levels of performance in recommender systems. 
The design of interaction between dense/sparse features is the key driver to optimizing the recommender models. 
The advancement of feature interactions incorporates improved prior knowledge into the recommender systems, enhancing the underlying user-item relationships and benefits personalization benchmarks such as Click-Through Rate (CTR) prediction.
In past practices, there have been several advancements in feature interactions, such as collaborative filtering~\cite{barkan2016item2vec,wang2016structural,he2017neural,zhang2017collaborative}.
With the rise of Deep Learning (DL), factorization machines~\cite{guo2017deepfm,lian2018xdeepfm}, DotProduct~\cite{cheng2016wide,naumov2019deep}, deep crossing~\cite{wang2021dcn}, and self-attention~\cite{song2019autoint}.
The stack (e.g., DHEN~\cite{zhang2022dhen}) and the combination (e.g., Mixture of Experts~\cite{masoudnia2014mixture,balog2006formal}) create versatile interaction types with varying orders, types, and dense/sparse input sources.

\begin{figure}[t]
    \begin{center}
    \includegraphics[width=0.75\linewidth]{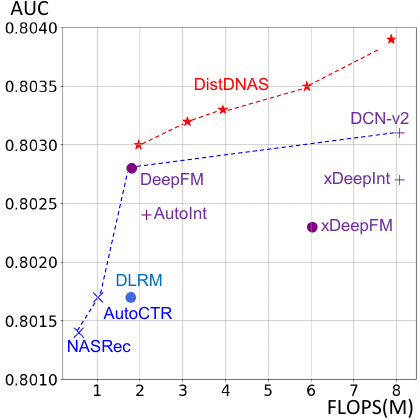}
    \vspace{-1em}
    \caption{Model AUC versus FLOPs on Criteo Terabyte. DistDNAS unlocks 0.001 AUC compared to state-of-the-art recommender models.}
    \label{fig:distdnas_sota}
    \end{center}
    \vspace{-2em}
\end{figure}

The recent thriving of Automated Machine Learning (AutoML) democratized the design of feature interactions and exceeded human performance in various domains, such as feature selection~\cite{liu2020autofis} and Neural Architecture Search~\cite{song2020towards,zhang2022nasrec,krishna2021differentiable}.
Remarkably, NASRec~\cite{zhang2022nasrec} employs a supernet to represent the search space for recommender models and achieves state-of-the-art results on small-scale CTR benchmarks.
However, optimizing feature interactions through manual design/search for large-scale CTR prediction has two challenges.
First, designing/searching for the optimal feature interactions requires
extensive wall-clock time, as the design/search sequentially iterates high data volume in production to obtain a good solution in feature interaction.
This raises obstacles to ensuring the freshness of feature interaction on the latest data, thus potentially harming production quality and causing model staleness.
Thus, an efficient search methodology is needed to explore the selection of the feature interaction and scale with the data volume growth.
Second, fusing versatile feature interactions introduces potential conflicts and redundancy in recommender models.
For example, we observe that loss divergence issues can be directly attributed to a complex interaction module (e.g., xDeepInt) with alternative feature interaction modules (e.g., CrossNet, DLRM) when training on a large dataset.
This creates a sub-optimality of performance-efficiency trade-offs in service. 
As the relationship between model size and model performance in recommender systems has not been exploited yet, it is uncertain whether fusing multiple orders and types of feature interactions into a single architecture can benefit recommender models or whether it is possible to harvest the improvement from different feature interactions fully. 

In this paper, we address the above challenges and present an efficient AutoML system, \textbf{Dist}ributed \textbf{D}ifferentiable \textbf{N}eural \textbf{A}rchitecture \textbf{S}earch (DistDNAS), to craft efficient feature interactions in a few hours. 
DistDNAS follows the setting of the supernet-based approach in NASRec~\cite{zhang2022nasrec} and applies Differentiable Neural Architecture Search ~\cite{liu2018darts} (DNAS) to learn the structure of a feature interaction. 
In DistDNAS, a feature interaction is built with multiple choice blocks. 
Each choice block represents a linear combination of feature interaction modules (e.g., Linear, CrossNet~\cite{wang2021dcn}, etc.).
DistDNAS presents two robust techniques to improve design and model efficiency during feature interaction search.
For search efficiency, DistDNAS distributes DNAS on each training day and averages the learned weighting to derive the best combinatorial choice of interaction modules.
Without ad-hoc optimizations such as embedding table sharding and communication between multiple devices, DistDNAS exhibits better scalability with a large volume of training data, achieving $\sim25\times$ speedup on 1TB Criteo Terabyte benchmark and reducing the search cost from 2 days to 2 hours.
For serving efficiency, DistDNAS calculates the cost importance of each interaction module and incorporates a differentiable cost-aware regularization loss to penalize cost-expensive interaction modules.
As cost-aware regularization loss removes unnecessary interaction modules within the supernet, DistDNAS alleviates potential conflict and harvests performance improvement. 

We evaluate DistDNAS on the 1TB Criteo Terabyte dataset using AUC, LogLoss, and Normalized Entropy (NE)~\cite{he2014practical} as evaluation metrics.
DistDNAS removes redundant interaction modules without human intervention and discovers efficient feature interactions, unlocking 0.001 higher AUC and/or 60\% fewer FLOPs in the discovered models. The optimization in AUC and FLOPs brings state-of-the-art models, see Figure \ref{fig:distdnas_sota}. We summarize our contributions as follows.
\begin{itemize}[noitemsep,leftmargin=*]
    \item We analyze the search and serving efficiency challenges when designing feature interactions on large-scale CTR recommender benchmarks.
    \item We propose DistDNAS, an AutoML system, to tackle the efficiency challenge in feature interaction design. DistDNAS distributes the search over multiple data splits and averages the learned architecture on each data split for search efficiency. In addition, DistDNAS incorporates a cost-aware regularization into the search to enhance the serving efficiency of discovered feature interactions.
    \item Our empirical evaluations demonstrate that DistDNAS significantly pushes the Pareto frontier of state-of-the-art CTR models.
\end{itemize}
\section{Related Work}

\noindent \textbf{Feature Interactions in Recommender Systems.}
The feature interactions within recommender systems such as CTR prediction have been thoroughly investigated in various approaches, such as Logistic Regression~\cite{richardson2007predicting}, and Gradient-Boosting Decision Trees~\cite{he2014practical}.
Recent approaches apply deep learning based interaction~\cite{zhang2019deep} to enhance end-to-end modeling experience by innovating Wide \& Deep Neural Networks~\cite{cheng2016wide}, Deep Crossing~\cite{wang2017deep}, Factorization Machines~\cite{guo2017deepfm,lian2018xdeepfm}, DotProduct~\cite{naumov2019deep} and gating mechanism~\cite{wang2017deep, wang2021dcn}, ensemble of feature interactions~\cite{zhang2022dhen}, feature-wise multiplications~\cite{wang2021masknet}, and sparsifications~\cite{deng2021deeplight}.
In addition, these works do not fully consider the impact of fusing different types of feature interactions, such as the potential redundancy, conflict, and performance enhancement induced by a variety of feature interactions.
DistDNAS constructs a supernet to explore different orders and types of interaction modules and distributes differentiable search to advance search efficiency.

\noindent \textbf{Cost-aware Neural Architecture Search.}     
Neural Architecture Search (NAS) automates the design of Deep Neural Network (DNN) in various applications: the popularity of NAS is consistently growing in brewing Computer Vision~\cite{zoph2018learning,liu2018darts,wen2020neural,cai2019once}, Natural Language Processing~\cite{so2019evolved,wang2020hat}, and Recommender Systems~\cite{song2020towards,gao2021progressive,krishna2021differentiable,zhang2022nasrec}.
Tremendous efforts are made to advance the performance of discovered architectures to brew a state-of-the-art model.
Despite the improvement in search/evaluation algorithms, existing NAS algorithms overlook the opportunity to harvest performance improvements by addressing potential conflict and redundancy in feature interaction modules.
DistDNAS regularizes the cost of the searched feature interactions and prunes unnecessary interaction modules as building blocks, yielding better FLOPs-LogLoss trade-offs on CTR benchmarks.
\begin{figure*}[t]
    \begin{center}
    \includegraphics[width=0.75\linewidth]{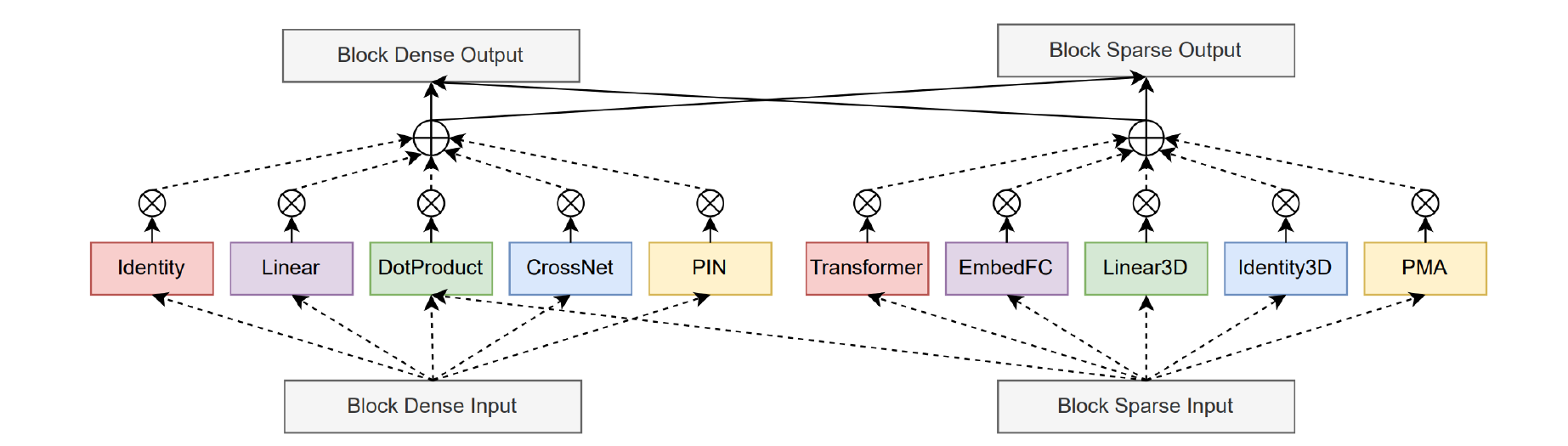}
    \vspace{-1em}
    \caption{Feature interaction search space for each choice block in DistDNAS. Here, a dashed line denotes a searchable feature interaction in DistDNAS, and $\otimes$ denotes the mixing of different feature interaction modules.}
    \label{fig:dnas_search_space}
    \end{center}
\vspace{-2em}
\end{figure*}

\section{Differentiable Feature Interaction Space}
Supernet is a natural fusion that incorporates feature interaction modules. DistDNAS emphasizes search on feature interaction modules and simplifies the search space from NASRec~\cite{zhang2022nasrec}. 
A supernet in DistDNAS contains multiple choice blocks, with a fixed connection between them and a fully enabled dense-to-sparse/sparse-to-dense merger within all choice blocks. 
Unlike NASRec~\cite{zhang2022nasrec}, which solely selects the optimal interaction module within each choice block, DistDNAS can choose an arbitrary number of interaction modules within each choice block and use differentiable bi-level optimization~\cite{liu2018darts} to determine the best selection.
This allows flexibility in fusing varying feature interactions and obtaining the best combination with enhanced search efficiency.
We present the details of feature interaction modules as follows.

\subsection{Feature Interaction Modules}
Feature interaction modules connect dense 2D inputs with sparse 3D inputs to learn valuable representations on user modeling.
In a recommender system, a dense input is a 2D tensor generated from either raw dense features or dense interaction modules, such as a fully connected layer.
A sparse input is a 3D tensor of sparse embeddings generated either by raw sparse/categorical features or by sparse interaction modules, such as a self-attention layer.
We define a dense input as $X_{d} \in \mathbb{R}^{B \times dim_d}$ and a sparse input $X_{s} \in \mathbb{R}^{B \times N_s \times dim_s}$. 
Here, B denotes the batch size, $dim_d$/$dim_s$ denotes the dimension of the dense/sparse input, and $N_s$ denotes the number of inputs in the sparse input.

We collect a set of simple feature interaction modules from the existing literature, as demonstrated in Figure \ref{fig:dnas_search_space}.
A dense interaction module produces a dense output given input features, and a sparse interaction module produces a sparse output given input features.
These interaction modules can cover a set of state-of-the-art CTR models, such as DLRM~\cite{naumov2019deep}, DeepFM~\cite{wang2017deep}, xDeepInt~\cite{yan2023xdeepint}, DCN-v2~\cite{wang2021dcn}, and AutoInt~\cite{song2019autoint}.
\begin{itemize}[noitemsep,leftmargin=*]
    \item \textbf{Identity/Identity3D} is a dense/sparse interaction module that carries an identity transformation on dense/sparse input.

    \item  \textbf{Linear/Linear3D} is a dense/sparse interaction module that applies on 2D/3D dense inputs.

    \item \textbf{DotProduct}\cite{cheng2016wide,naumov2019deep} is a dense interaction module that computes pairwise inner products of dense and sparse inputs.

    \item \textbf{CrossNet}~\cite{wang2021dcn}/\textbf{PIN}~\cite{yan2023xdeepint} is a dense interaction module with gate inputs from various sources. 

    \item \textbf{Transformer}~\cite{vaswani2017attention} is a sparse interaction module that utilizes the multihead attention mechanism to learn the weighting of different sparse inputs. 
    The queries, keys, and values of a Transformer layer are identical.
    
    \item \textbf{Embedded Fully-Connected (EmbedFC)} is a sparse interaction module that applies a linear operation along the sparse (middle) dimension.

    \item \textbf{Pooling by Multihead Attention (PMA)}~\cite{lee2019set}  is a sparse interaction module that forms attention between seed vectors and sparse features.
\end{itemize}

If the dimensions of inputs do not match, a proper linear projection (e.g., linear) will be applied within all dense/sparse interaction modules. The feature interaction search space contains versatile dense/sparse interaction modules.

\begin{figure*}[t]
    \begin{center}
    \includegraphics[width=0.75\linewidth]{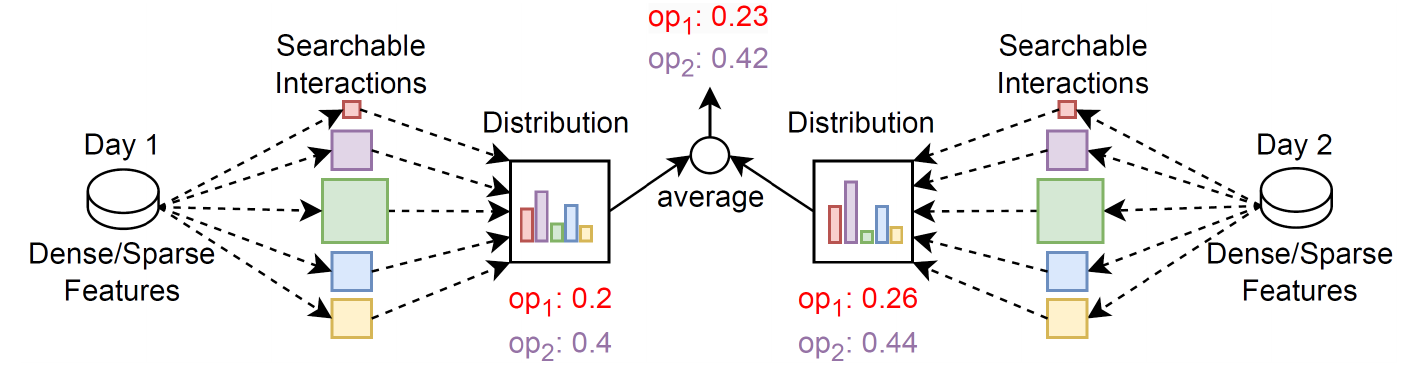}
    \vspace{-1em}
    \caption{Overview of DistDNAS methodology. Here, dashed lines denote searchable interaction modules, and the size of interaction modules indicates the cost penalty applied to each interaction module for serving efficiency.}
    \label{fig:distributed_dnas}
    \vspace{-2em}
    \end{center}
\end{figure*}
\subsection{Differentiable Supernet}
In DistDNAS, a differentiable supernet contains $N$ choice blocks with a rich collection of feature interactions.
Each choice block employs a set of dense/sparse interactions $op^{(i)}=\{op_{d}^{(i)}, op_{s}^{(i)}\}$ to take dense/sparse inputs $X^{(i)}_{d}$/$X^{(i)}_{s}$ and learn useful representations. 
Each choice block contains $|op_{d}|=5$ dense feature interactions and $|op_{s}|=5$ sparse feature interactions. 
Each choice block receives input from previous choice blocks and produces a dense output $Y_d^{(i)}$ and a sparse output $Y_{s}^{(i)}$. 
In block-wise feature aggregation, each choice block concatenates the dense/sparse output from the previous 2/1 choice blocks as dense/sparse block input. 
For dense inputs in previous blocks, the concatenation occurs in the last feature dimension. 
For sparse inputs in previous blocks, concatenation occurs in the middle dimension to aggregate different sparse inputs.
We present the mixing of different feature interaction modules in the following context.

\noindent \textbf{Continuous Relaxation of Feature Interactions.}
Within a single choice block in the differentiable supernet, we depict the mixing of candidate feature interaction modules in Figure \ref{fig:dnas_search_space}.
We parameterize the weighting of each dense/sparse/dense-sparse interaction using architecture weights. 
In choice block $i$, we use $\alpha^{(i)}=\{\alpha_{1}^{(i)}, \alpha_{2}^{(i)}, ..., \alpha_{|op_{d}|}^{(i)}\}$ to represent the weighting of a dense interaction module and parameterize the selection of dense interaction modules with architecture weight $\mathrm{A}=\{\alpha^{(1)}, \alpha^{(2)}, ..., \alpha^{(N)}\}$.
Similarly, we parameterize the selection of sparse interaction modules with architecture weight $\mathrm{B}=\{ \beta^{(1)}, \beta^{(2)}, ..., \beta^{(N)} \}$. We employ the Gumbel Softmax~\cite{jang2016categorical} trick to allow a smoother sampling from categorical distribution on dense/sparse inputs as follows: 
\begin{equation}
Y_{d}^{(i)} = \sum_{j=1}^{|op_d|}\frac{\exp(\frac{\log \alpha_{j}^{(i)} + g_{j}}{\lambda})}{\sum_{k}^{|op_d|}\exp(\frac{\log \alpha_{k}^{(i)} + g_{k}}{\lambda})} op_{j}(X_{d}^{i},X_{d}^{i-1}),
    \label{eq:dense_block_out_dnas}
\end{equation}
\vspace{-0.5em}
\begin{equation}
    Y_{s}^{(i)} = \sum_{j=1}^{|op_s|} \frac{\exp{(\frac{\log \beta_{j}^{(i)} + g_{j}}{\lambda}})}{\sum_{k}^{|op_s|} \exp(\frac{\log {\beta_k}^{(i)} + g_{k}}{\lambda})} op_{j}(X_{s}^{i}, X_{s}^{i-1})
    \label{eq:sparse_block_out_dnas}.
\end{equation}

Here, $g_{j}$ and $g_{k}$ are sampled from the Gumbel distribution, and $\lambda$ is the temperature. As a result, a candidate architecture $\mathcal{C}$ can be represented as a tuple of dense/sparse feature interaction: $\mathcal{C}=(\mathrm{A}, \mathrm{B})$. Within DistDNAS, our goal is to perform a differentiable search and obtain the optimal architecture $C^{*}$ that contains the weighting of dense/sparse interaction modules. 

\noindent \textbf{Discretization.} Discretization converts the weighting in the optimal architecture to standalone models to serve CTR applications. 
Past DNAS practices~\cite{liu2018darts} typically select the top k modules for each choice block, brewing a building cell containing a fixed number of modules in all parts of the network.
In recommender models, we discretize each choice block by a fixed threshold $\theta$ to determine useful interaction modules. For example, in the choice block $i$, we discretize the weight $\alpha^{(i)}$ to obtain the discretized dense interaction module $\hat{\alpha}^{(i)}$ as follows:
$$
\hat{\alpha}^{(i)}_{j} =
\begin{cases}
1, if    \hat{\alpha}^{(i)}_{j} \geq \theta, \\
0, otherwise.\\
\end{cases}
$$
We typically set threshold $\theta=1/|op|$ (i.e., 0.2 for dense/sparse module search) for each choice block, or use a slightly larger value (e.g., 0.25) to remove more redundancy.
There are a few advantages of adopting threshold-based discretization in recommender models. 
First, using a threshold $\theta$ is a clearer criterion to distinguish important/unimportant interaction modules within each choice block.
Second, since a recommender model contains multiple choice blocks with different hierarchies, levels, and dense/sparse input sources, there is a need for varying numbers of dense/sparse interactions to maximize the representation capacity within each module.

\section{Towards Efficiency in DNAS}
Search efficiency and serving efficiency are two major considerations in deploying DNAS algorithms in large-scale CTR datasets.
In this section, we first revisit DNAS and address the efficiency bottleneck via a distributed search mechanism. Then, we propose our solution to reduce the service cost of feature interaction via a cost-aware regularization approach. 
Figure \ref{fig:distributed_dnas} provides the core methodology of DistDNAS.

\label{sec4:dist}
\subsection{Revisiting DNAS on Recommender Systems}
\begin{figure}[t]
    \begin{center}
    \includegraphics[width=0.75\linewidth]{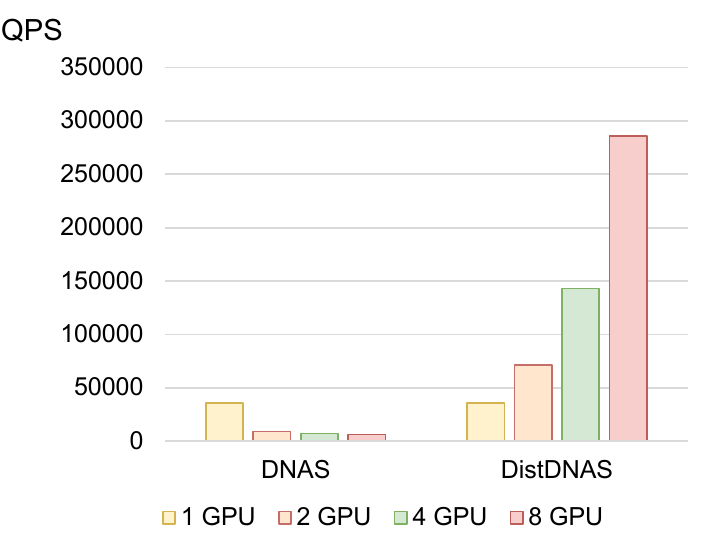}
    \vspace{-1.5em}
    \caption{QPS comparison between DistDNAS and DNAS.}
    \label{fig:dnas_qps}
    \end{center}
    \vspace{-2em}
\end{figure}

A Click-Through Rate (CTR) prediction task usually contains multiple days of training data. In recommender systems, we typically use a few days of data (i.e., day 1 to day $T$) as the training source and evaluate the trained model based on its CTR prediction over subsequent days. DNAS carries bilevel optimization to find the optimal candidate architecture $C^{*}=(A^{*}, B^{*})$ as follows:
\begin{equation}
    (\mathrm{A}^{*}, \mathrm{B}^{*}) = \argmin_{\mathrm{A}, \mathrm{B}} LogLoss_{x \sim D}[x; w^{*}(\mathrm{A}, \mathrm{B}), \mathrm{A}, \mathrm{B}].
\end{equation}
Here, $BCE$ denotes binary cross-entropy, $D=(D_1, ..., D_{T})$ indicates the training data from day 1 to day $T$, $w$ indicates the weight parameters within the DNN architecture, and $t$ indicates a certain day of data. 
The previous DNAS workflow must iterate over $T$ days of data, with a significant search cost. More specifically, the large search cost originates from the following considerations in search efficiency and scalability:
\begin{itemize}[noitemsep,leftmargin=*]
    \item Sequentially iterating over $T$ days of data requires $T$ times the search cost of DNAS on a single day of data. This creates challenges for model freshness in production environments where $T$ can be extremely large. 
    \item Deploying the search over multiple devices may suffer from poor scalability due to communication. For example, the forward/backward process must shift from model to data parallelism when offloading tensors from embedding table shards to feature interaction modules.
    \item Within our implementation on Criteo Terabyte, the throughput on multiple NVIDIA A5000 GPUs is lower than the throughput on a single GPU during the search, as demonstrated in the Queries-Per-Second (QPS) analysis in Figure \ref{fig:dnas_qps}. 
    Thus, realizing good scalability in the growth of computing devices in DNAS is challenging.
\end{itemize}
 The above considerations point to a distributed version of DNAS, where we partition the training data, launch a DNAS procedure on each day of training data, and average the results to derive the final architecture.
 We hereby propose a DistDNAS search with the following bilevel optimization.
\begin{equation}
    (\mathrm{A}^{*}, \mathrm{B}^{*})^{(t)} = \argmin_{\mathrm{A}, \mathrm{B}} LogLoss_{x \sim D_{t}}[x; w^{*}(\mathrm{A}, \mathrm{B}), \mathrm{A}, \mathrm{B}],
\end{equation}
such that
\begin{equation}
    w^{*}(\mathrm{A}, \mathrm{B})^{(t)} = \argmin_{w} LogLoss_{x \sim D_{t}}[x; w(\mathrm{A}, \mathrm{B}), \mathrm{A}, \mathrm{B}].
\end{equation}

Here, $t \in \{1, 2, ..,, T\}$ indicates a certain day of training data. 
DistDNAS aggregates the learned weights on each day to retrieve the final architecture coefficients with a simple averaging aggregator as follows:
\begin{equation}
\small
    (\mathrm{A}^{*}, \mathrm{B}^{*})^{(*)} = \sum_{t=1}^{T} \frac{1}{T}(\mathrm{A}^{*}, \mathrm{B}^{*})^{(t)}
\end{equation}
The simple averaging scheme incorporates the statistics from each day of data to obtain the learned architecture weights. 
In addition, DistDNAS can be asynchronously paralleled on different computing devices, accelerating search scalability and reducing the total wall-clock run-time in recommender systems. Figure \ref{fig:dnas_qps} compares DistDNAS versus DNAS on 1-8 NVIDIA A5000 GPUs, with 4K batch size. Due to communication savings with 1-GPU training, DistDNAS benefits from significantly lower search costs.

\begin{figure}[b]
    \begin{center}
    \vspace{-2em}
    \includegraphics[width=0.75\linewidth]{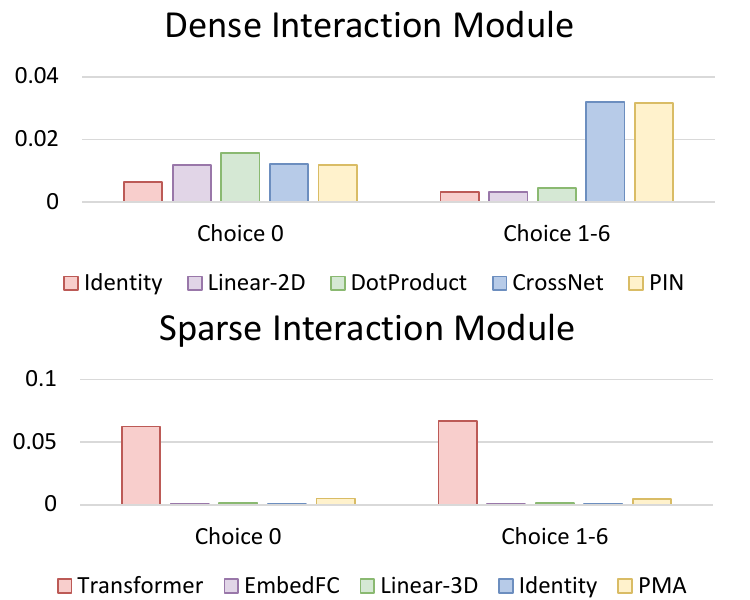}
    \vspace{-1em}
    \caption{Normalized cost importance in a 7-block supernet.}
    \label{fig:flops_feature_interactions}    
    \end{center}
\end{figure}

\subsection{Cost-aware Regularization}
Serving the cost of feature interactions, e.g., FLoating-point OPerations (FLOPs), is critical in recommender systems. A lower servicing cost indicates a shorter response time to process a user query request. As a result, optimizing the cost of recommender models is as important as optimizing their performance in the production environment.

In DNAS, we measure the cost by combining training FLOPs and inference latency.
The cost of a feature interaction in discovery is dependent on the weights of the learned architecture $C^{*}=(A^{*}, B^{*})$. An intuition to optimize the feature interaction module is rewarding cost-efficient operators (e.g., Linear, Identity) while penalizing cost-inefficient operators (e.g., Transformer, CrossNet) during differentiable search.
Motivated by this, we introduce a differentiable cost regularizer to penalize large models in discovery.
The cost regularizer adds an additional regularization term $R$ to the loss function during DNAS to induce cost-effective feature interactions in discovery.

We use $j$ to represent an index of a feature interaction module in $\mathcal{C}$, for example, the index of a dense interaction module.
We first sample a few pairs of architecture and cost metrics from the DistDNAS search space and create a cost mapping $cost: \mathrm{C} \to \mathrm{R}$ to model the relationship between feature interactions and FLOPs.
Then, we use the permutation importance~\cite{breiman2001random} to obtain the importance of offline cost $s_{j}$ in the cost mapping $cost$, illustrating the offline FLOPs importance of an interaction module $i$. 
Finally, we formulate a cost-aware loss and incorporate it to regularize all interaction modules: $R(\mathrm{A}, \mathrm{B}) = \gamma \sum_{op^{j} \in (\mathrm{A}, \mathrm{B})}s_{j}$.
Here, $\gamma$ is an adjustable coefficient to control the strength of cost-aware regularization. With cost-aware regularization, the final architecture of DNAS with cost-aware loss searched on a single day $t$ can be formulated as follows.
\begin{equation}
    \small
    (\mathrm{A}^{*}, \mathrm{B}^{*})^{(t)} = \argmin_{\mathrm{A}, \mathrm{B}} LogLoss_{x \sim D_{t}}[x; w^{*}(\mathrm{A}, \mathrm{B}), \mathrm{A}, \mathrm{B}] + R(\mathrm{A}, \mathrm{B}),
\end{equation}
such that 
\begin{equation}
\small
    w^{*}(\mathrm{A}, \mathrm{B})^{(t)} = \argmin_{w} LogLoss_{x \sim D_{t}}[x; w(\mathrm{A}, \mathrm{B}), \mathrm{A}, \mathrm{B}] + R(\mathrm{A}, \mathrm{B}).
\end{equation}

As the feature interaction search space adopts a fixed connectivity and dimension configuration during the search, the cost importance of different interaction modules is unique in the first choice block, and identical across all other choice blocks.
We demonstrate the normalized cost importance of each interaction module in Figure \ref{fig:flops_feature_interactions}. 
Among all interaction modules, the DotProduct contributes to a significant amount of FLOPs consumption by integrating dense and/or sparse features. Except for Transformer, sparse interaction modules contribute significantly fewer serving costs compared to their dense counterparts.
Thus, despite the strong empirical performance of Transformer models, recommender models choose Transformer sparingly to build an efficient feature interaction.

\begin{table*}[t]
    \begin{center}
    \caption{Performance of the best discovered DistDNAS model on 1TB Criteo Terabyte.}
    \vspace{-1em}
    \scalebox{1.0}{
    \begin{tabular}{|c|c|c|c|c|c|c|}
    \hline
    \textbf{Model} & \textbf{FLOPS(M)} & \textbf{Params(M)} & \textbf{NE (\%)} $\downarrow$ & \textbf{Relative NE (\%)} $\downarrow$ & \textbf{AUC} $\uparrow$  & \textbf{LogLoss} $\downarrow$   \\
    \hline \hline
    DLRM~\cite{naumov2019deep} & 1.79 & 453.60 & 0.8462 & 0.0 & 0.8017 & 0.12432  \\
    DeepFM~\cite{guo2017deepfm} & 1.81 & 453.64 & 0.845 & -0.14 & 0.8028 & 0.12413 \\
    xDeepFM~\cite{lian2018xdeepfm} & 6.03 & 454.14 & 0.846 & -0.02 & 0.8023 & 0.12429 \\
    AutoInt~\cite{song2019autoint} & 2.16 & 453.80 & 0.8455 & -0.08 & 0.8024 & 0.12421 \\
    DCN-v2~\cite{wang2021dcn} & 8.08 & 459.91 & 0.845 & -0.14 & 0.8031 & 0.12413  \\
    xDeepInt~\cite{yan2023xdeepint} & 8.08 & 459.91 & 0.8455 & -0.08 & 0.8027 & 0.12421 \\
    \hline
    NASRec-tiny~\cite{zhang2022nasrec} & 0.57 & 452.47 & 0.8463 & 0.01 &  0.8014 & 0.12437 \\
    AutoCTR-tiny~\cite{song2020towards} & 1.02 & 452.78 & 0.8460 & -0.02 & 0.8017 & 0.12429 \\
    \multirow{2}{*}{DistDNAS} & 1.97 & 453.62 & \textbf{0.8448} & \textbf{-0.17} & 0.8030 & \textbf{0.12410} \\
    & $3.11^{*}$ & 454.70 & \textbf{0.8444} & \textbf{-0.21} & \textbf{0.8032} & \textbf{0.12405} \\
    DistDNAS (M=2) & 3.94 & 455.48 & \textbf{0.8448} & \textbf{-0.17} & \textbf{0.8033} & \textbf{0.12410} \\
    DistDNAS (M=3) & 5.90 & 457.31 & \textbf{0.8440} & \textbf{-0.26} & \textbf{0.8035} & \textbf{0.12399} \\
    DistDNAS (M=4) & 7.87 & 459.14 & \textbf{0.8438} & \textbf{-0.29} & \textbf{0.8039} & \textbf{0.12395} \\
    \hline
    \end{tabular}
    \label{tab:criteo_terabyte_tab} 
    }
    \end{center}
    \vspace{-2em}
\end{table*}
\section{Experiments}
We thoroughly examine DistDNAS on Criteo Terabyte.
First, we introduce the experiment settings of DistDNAS, which produce efficient feature interactions.
Then, we compare the performance of models crafted by DistDNAS versus a series of metrics with strong hand-crafted/AutoML baselines.


\subsection{Experiment Setup} 
We illustrate the key components of our experiment setup and elaborate on the detailed configuration.

\noindent \textbf{Training Dataset.}
Criteo Terabyte contains 4B training data on 24 different days. Each data item contains 13 integer features and 26 categorical features.
Each day of data on Criteo Terabyte contains $\sim$ 0.2B data.
During the DistDNAS search, we use data from day 1 to day 22 to learn architecture the optimal architecture weights: $C^{*}=(A^{*}, B^{*})$. During the evaluation, we use the data from day 1 to day 23 as a training dataset and use \textbf{day 24} as a holdout testing dataset. We perform inter-day data shuffling during training yet iterate over data from day 1 to day 23 sequentially.

\noindent \textbf{Data Preprocessing.} We do not apply special preprocessing to dense features except for normalization. We cap the maximum embedding table size to 5M for sparse embedding tables and use an embedding dimension of 16 to obtain each sparse feature. Thus, each model contains $\sim$450M parameters in the embedding table.

\noindent \textbf{Optimization.} 
We train all models from scratch without inheriting knowledge from other sources, such as pre-trained models or knowledge distillation.
For sparse parameters, we utilize Adagrad with a learning rate of 0.04. For dense parameters, we use Adam with a learning rate of 0.001. 
No weight decay is performed.
During training, we use a fixed batch size of 8192 and a fixed learning rate after the initial warm-up. 

\noindent \textbf{Architecture Search.}
Our supernet contains $N=7$ choice blocks during the search.
We choose $\gamma$=0.004 for cost-aware regularization.
During the search, we linearly warm up the learning rate from 0 to maximum with 10K warm-up steps and use a batch size of 8K to learn the architecture weights while optimizing the DNAS supernet.
During discretization,  we use $\theta$=0.20 as the discretization threshold for DistDNAS marked with $*$, and use $\theta=0.25$ to other feature interactions created by DistDNAS, in Table \ref{tab:criteo_terabyte_tab}, 
As most baseline models are larger, we naively stack $M$ copies of feature interactions in parallel to match the FLOPs of large models, such as DCN-v2~\cite{wang2021dcn} and xDeepInt~\cite{yan2023xdeepint}.

\noindent  \textbf{Training.} To ensure a fair comparison and better demonstrate the strength of the discovered models, we employ no hyperparameter tuning for all models.
We linearly warm up the learning rate from 0 to maximum using the first 2 days of training data.
We use single-pass training to prevent overfitting and iterate the whole training dataset only once.

\noindent \textbf{Baselines.} We select the popular hand-crafted design choice of CTR models from the existing literature to serve as baselines, i.e., DLRM~\cite{naumov2019deep}, DeepFM~\cite{guo2017deepfm}, xDeepFM~\cite{lian2018xdeepfm}, AutoInt~\cite{song2019autoint}, DCN-v2~\cite{wang2021dcn} and xDeepInt~\cite{yan2023xdeepint}. 
We also incorporate the best models from the NAS literature: AutoCTR~\cite{song2020towards} and NASRec~\cite{zhang2022nasrec} to serve as baselines and use the best model discovered for Criteo Kaggle.
Without further specification, 
all hand-crafted or AutoML baselines use $dim_{s}=16$ as the embedding dimension.
All hand-crafted or AutoML baselines use 512 or $dim_{d}=256$ units in the MLP layer, including 1 MLP layer in dense feature processing and 7 MLP layer in aggregating high-level dense/sparse features.
All AutoML models use $N_{s}=16$ for sparse interaction modules.
This ensures a fair comparison between hand-crafted and AutoML models, as the widest part in hand-crafted/AutoML models does not exceed 512.
All hand-crafted feature interactions (e.g., CrossNet) are stacked 7 times to match $N=7$ blocks in the AutoML supernet, as NASRec, AutoCTR, and proposed DistDNAS employ $N=7$ blocks for feature interaction. We name the derived NAS baselines \textbf{NASRec (tiny)} and \textbf{AutoCTR (tiny)}. We implement all baseline feature interactions based on open-source code and/or paper demonstration. 

\subsection{Evaluation on Criteo Terabyte}
We use DistDNAS to represent the performance of the best models discovered by DistDNAS and compare performance against a series of cost metrics such as FLOPs and parameters. 
We use AUC, Normalized Entropy (NE)~\cite{he2014practical}, and LogLoss as evaluation metrics to measure model performance.
We also calculate the testing NE of each model relative to DLRM and demonstrate relative performance. Note that relative NE is equivalent to relative LogLoss on the same testing day of data. 
Table \ref{tab:criteo_terabyte_tab} summarizes our evaluation of DistDNAS.
Here, $M$ indicates the number of parallel stackings we apply on DistDNAS to match the FLOPs of baseline models.

Upon transferring to large datasets, previous AutoML models~\cite{song2020towards,zhang2022nasrec} searched on smaller datasets are less competitive when applied to large-scale Criteo Terabyte. This is due to sub-optimal architecture transferability from the source dataset (i.e., Criteo Kaggle) to the target dataset (i.e., Criteo Terabyte). Among all baseline models, DCN-v2 achieves state-of-the-art performance on Criteo Terabyte with the lowest LogLoss/NE and highest AUC. 
DistDNAS shows remarkable model efficiency by establishing a new Pareto frontier on AUC/NE versus FLOPs.
With a discretization threshold of 0.25, DistDNAS outperforms tiny baseline models such as DLRM and DeepFM and unlocks at least 0.02\% AUC/NE with on-par FLOPs complexity.
With a discretization threshold of 0.2, DistDNAS achieves better AUC/NE as state-of-the-art DCN-v2 models, yet with a reduction of over 60\% FLOPS.
By naively stacking more blocks in parallel, DistDNAS outperforms DCN-v2 by 0.001 AUC and achieves state-of-the-art.

\section{Discussion}
In this section, we conduct ablation studies and analyze various confounding factors within DistDNAS.
\begin{figure*}[t]
    \begin{center}
    \includegraphics[width=0.7\linewidth]{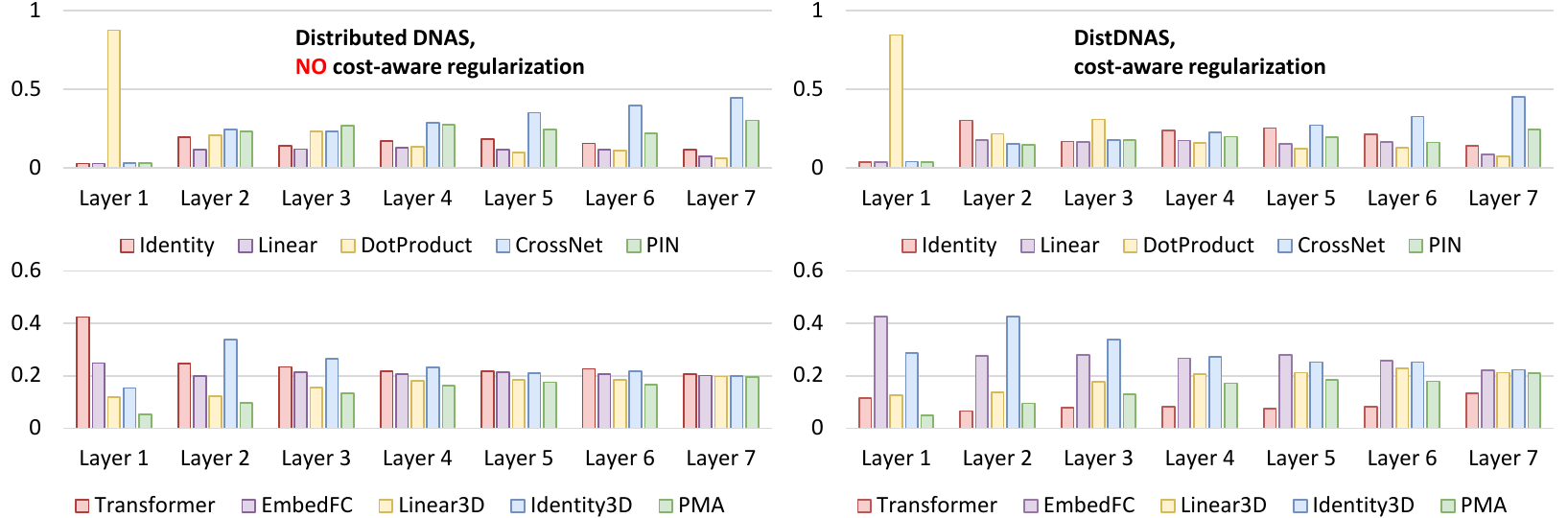}
    \vspace{-1em}
    \caption{Comparison of learned architecture weights under distributed DNAS versus DistDNAS.}
    \label{fig:cost_ablation}   
    \vspace{-2em}
    \end{center}
\end{figure*}
\subsection{DistDNAS Search Strategy}
We discuss the alternative choices to DistDNAS as follows.
\noindent \textbf{SuperNet} indicates a direct use of the DistDNAS supernet as a feature interaction module. No search is performed.

\noindent \textbf{Distributed DNAS} applies the DistDNAS search process in a distributed manner but does not involve cost-aware regularization. Both distributed DNAS and DistDNAS take 2 hours to complete on NVIDIA A5000 GPUs.

\noindent \textbf{One-shot DNAS} kicks off DNAS and iteratively over the entire search data set (that is, 22 days in Criteo Terabyte) to obtain the best architecture. Running a one-shot DNAS takes $\sim$ 50 GPU hours on an NVIDIA A5000 GPU.

\noindent \textbf{Fresh DNAS} only uses the most recent data in the training data set (that is, day 22 on Criteo Terabyte) and performs a search to learn the best architecture. This serves as a strong baseline due to the correlation between testing data set and the most recent data.

\noindent \textbf{DistDNAS} applies all the techniques proposed in this paper, including distributed search and cost-aware regularization. Figure \ref{fig:cost_ablation} compares the learned architecture weights in Distributed DNAS versus DistDNAS.
DistDNAS is more likely to preserve cost-efficient interaction modules like EmbedFC/Identity than distributed DNAS without cost-aware regularization.

We perform each of the searches above and evaluate different search strategies based on the following questions:
\begin{itemize}[noitemsep,leftmargin=*]
    \item \textbf{(Search Convergence)} Whether the search converges and produces effective feature interactions?
    \item \textbf{(Training Convergence)} Does the discovered feature interaction converge on a large-scale Criteo Terabyte benchmark with 24 days of training data?
    \item \textbf{(Testing Quality)} What is the quality of the interactions of the features discovered?
\end{itemize}

\begin{table}[b]
\begin{center}
    \vspace{-2em}
    \caption{Study of different DistDNAS search strategies.}
    \vspace{-1em}
    \begin{scalebox}{1.0}{
        \begin{tabular}{|c|c|c|c|}
        \hline
        \multirow{2}{*}{\textbf{Strategy}} & \textbf{Searching} & \textbf{Training} & \textbf{Testing}  \\
        & \textbf{Converge?} & \textbf{Converge?} & \textbf{FLOPs/NE} \\
        \hline
        \textbf{SuperNet} & N/A & No & N/A \\
        \textbf{One-shot DNAS} & No & No & N/A \\
        \textbf{Freshness DNAS} & Yes & No & N/A \\
        \textbf{Distributed DNAS} & Yes & Yes & 3.56M/0.8460 \\
        \textbf{DistDNAS} & Yes & Yes & 3.11M/0.8444 \\
        \hline
    \end{tabular}
    }    
    \end{scalebox}
    \label{tab:distdnas_strategy}    
\end{center}
\end{table}

Table \ref{tab:distdnas_strategy} summarizes a study of different search strategies for these questions. We have a few findings regarding the use of distributed search and cost-aware regularization.

A standalone supernet cannot converge when trained on Criteo Terabyte dataset, indicating that different feature interactions may have conflicts with each other.

On a large-scale dataset such as Criteo Terabyte, distributing DNAS over multiple-day splits and aggregating the learned weight architectures are critical to the convergence of search and training.
This is because in recommender systems, there might be an abrupt change in different user behaviors across/intra-days; thus, a standalone architecture learned on a single day may not be suitable to capture the knowledge and fit all user-item representations. Additionally, as NAS may overfit the target dataset, a standalone feature interaction searched on day $X$ may not be able to learn day $Y$ well and is likely to collapse due to changes in user behavior.

We also compare distributed DNAS with DistDNAS to demonstrate the importance of cost-aware regularization. 
Experimental evaluation demonstrates that FLOPS-regularization enhances the performance of searched feature interaction, removing the redundancy contained in the supernet. This observation provides another potential direction for recommender models to compress unnecessary characteristics and derive better recommender models, such as the usage of pruning.

\begin{figure}[t]
    \begin{center}
    \includegraphics[width=0.8\linewidth]{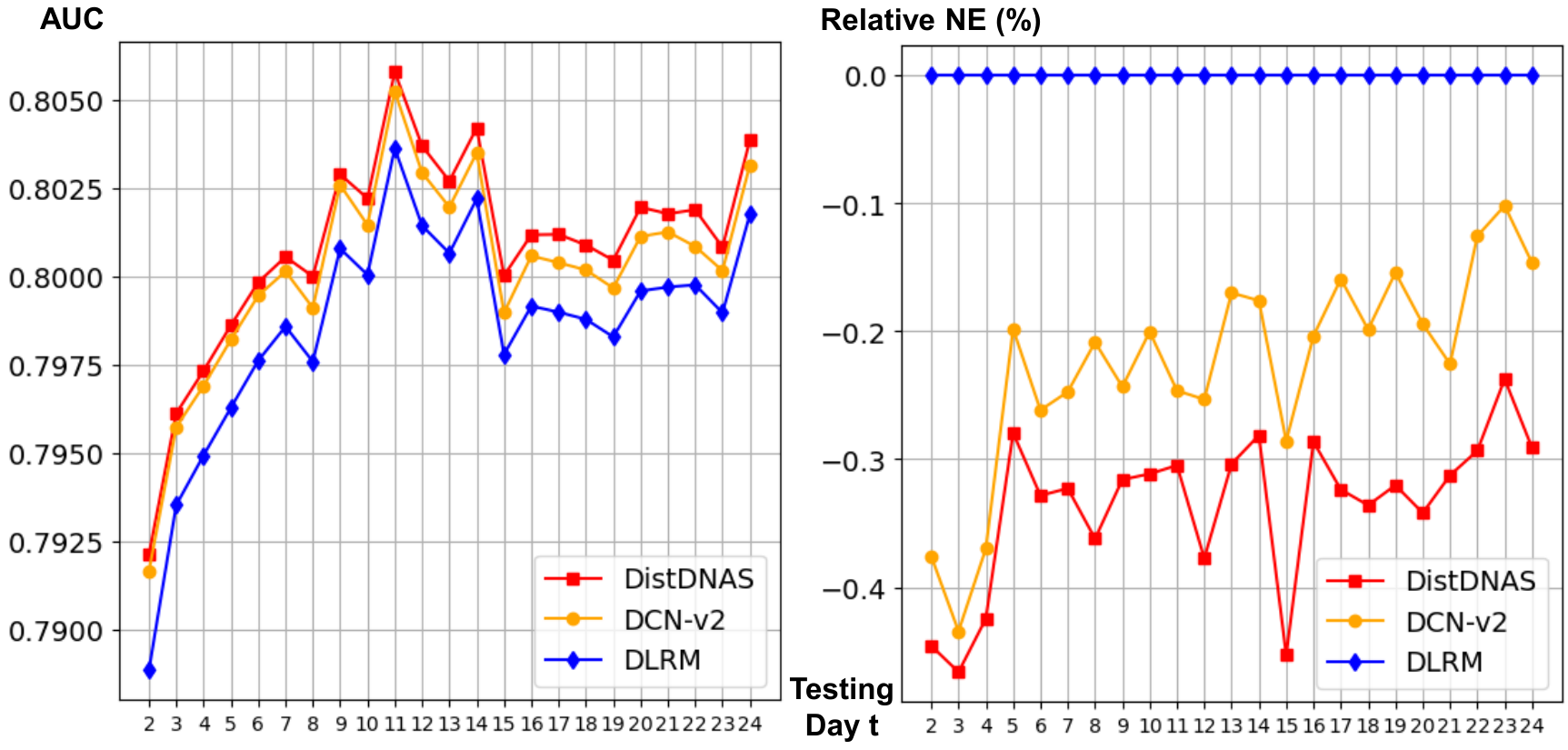}
    \vspace{-1em}
    \caption{Comparison of AUC and Relative NE under recurring training.}
    \label{fig:ablation_ne_recurring}    
    \end{center}
    \vspace{-2em}
\end{figure}

\subsection{Performance Analysis under Recurring Training Scenario}
Recurring training~\cite{he2014practical} is a common practice in recommender system applications. In recurring training, practitioners must regularly update the model on the latest data to gain fresh knowledge. Here, we simulate the scenario in recurring training to evaluate top-performing feature interactions (i.e., DistDNAS and DCN-v2) on different training/evaluation splits of Criteo Terabyte. We use day 1 to day $t$ as training dates and day $t+1$ as testing dates to report AUC and relative NE. 

We demonstrate the evaluation of recurring training on DistDNAS, DCN-v2, and DLRM (baseline) in Figure \ref{fig:ablation_ne_recurring}.
DistDNAS consistently outperforms the previous state-of-the-art DCN-v2 on all testing splits under recurring training. This indicates that DistDNAS successfully injects the implicit patterns contained within the large-scale dataset into the searched feature interaction.

\section{Conclusion}
In this article, we emphasize search efficiency and serving efficiency in the design of feature interactions through a differentiable supernet. 
We propose DistDNAS to explore the differentiable supernet containing various dense and sparse interaction modules.
We distribute the search on different days of training data to advance search scalability, reducing end-to-end search cost from 2 days to 2 hours with a 25$\times$ speed-up in scalability.
In addition, DistDNAS incorporates cost-aware regularization to remove potential conflicts and redundancies within feature interaction modules, yielding better performance and efficiency in searched architectures.

\noindent \textbf{Acknowledgment}. 
Yiran Chen’s work is partially supported by NSF-2330333, 2112562, 2120333, and ARO W911NF-23-2-0224. Feng Yan’s work is partially supported by NSF CAREER-2305491. Hai Li's work is partially supported by NSF-2148253.

\bibliographystyle{ieeetr}
\bibliography{references}

\end{document}